\documentclass[aps,pre,reprint,10pt,twocolumn,nofootinbib,showpacs]{revtex4-1}
\usepackage{graphicx}
\usepackage{amssymb}
\usepackage{amsmath}
\usepackage{amsfonts}
\usepackage{color}
\usepackage[latin1]{inputenc}
\usepackage{float}

\newcommand{\expct}[1]{\langle{#1}\rangle}

\begin{document}
\title{Height fluctuations in homoepitaxial thin film growth: A numerical study}

\author{I. S. S. Carrasco$^{(a,b)}$}
\email{ismael.carrasco@ufv.br}
\author{T. J. Oliveira$^{(a)}$}
\email{tiago@ufv.br}
\affiliation{$(a)$ Departamento de F\'isica, Universidade Federal de Vi\c cosa, 36570-900, Vi\c cosa, Minas Gerais, Brazil \\
$(b)$ Instituto de F\' isica, Universidade Federal Fluminense, Avenida Litor\^ anea s/n, 24210-340 Niter\' oi, Rio de Janeiro, Brazil}

\begin{abstract}
We report on the investigation of height distributions (HDs) and spatial covariances of two-dimensional surfaces obtained from extensive numerical simulations of the celebrated Clarke-Vvedensky (CV) model for homoepitaxial thin film growth. In this model, the effect of temperature, deposition flux, and strengths of atom-atom interactions are encoded in two parameters: the diffusion to deposition ratio $R=D/F$ and $\varepsilon$, which is related to the probability of an adatom ``breaking'' a lateral bond. We demonstrate that the HDs present a strong dependence on both $R$ and $\varepsilon$, and even after the deposition of $10^5$ monolayers (MLs) they are still far from the asymptotics in some cases. For instance, the temporal evolution of the HDs' skewness (kurtosis) displays a pronounced minimum (maximum), for small $R$ and $\varepsilon$, and only at long times it passes to increase (decrease) toward its asymptotic value. However, it is hard to determine whether they converge to a single value or different nonuniversal ones. For large $R$ and/or $\varepsilon$, on the other hand, these quantities clearly converge to the values expected for the Villain-Lai-Das Sarma (VLDS) universality class. A similar behavior is observed in the spatial covariances, but with weaker finite-time effects, so that rescaled curves of them collapse quite well with the one for the VLDS class at long times. Simulations of a model with limited mobility of particles, which captures some essential features of the CV model in the limit of irreversible aggregation ($\varepsilon=0$), reveal a similar scenario. Overall, these results point out that the study of fluctuations in homoepitaxial thin films' surfaces can be a very difficult task and shall be performed very carefully, once typical experimental films have $\lesssim 10^4$ MLs, so that their HDs and covariances can be in the realm of transient regimes.
\end{abstract}


\maketitle

\section{Introduction}
\label{secintro}

The list of applications of thin film deposition is very vast, underlying a number of recent important technological advances \cite{HB,orhing,Krugbook}. In most of the techniques and relevant conditions for thin film production, molecules coming from a vapor or a liquid phase adsorb onto the substrate or film surface (with a flux $F$ per adsorption site) and then passes to diffuse at the surface.
This thermally activated diffusion process may depend on the local environment of the adatoms, on the diffusion mechanism (e.g, by hopping or by exchange), on how adatoms interact at the surface, and so on \cite{orhing,Krugbook}. However, in most systems, it can be realistically modeled by simply considering the substrate and film surfaces as lattices of adsorption sites and diffusion occurring by the hopping of adatoms to nearest neighbor (NN) sites \cite{Evans2006}. 

In the case of homoepitaxial growth --- which is the one considered in this work ---, no distinction shall exist between the diffusion on the substrate or film surfaces. The hopping rate $\mathcal{H}$ is expected to follow an Arrhenius form $\mathcal{H} = \nu e^{-E/(k_B T)}$, where $\nu$ can be considered as a constant frequency and $E$ is the activation barrier for diffusion. Usually, in the modeling of these processes, $E$ is separated into a site-independent terrace diffusion barrier ($E_d$), and a contribution ($E_n$) which shall depends on the local environment of the adatom, i.e, the height configuration around the actual site and the target ones \cite{Evans2006,Krugbook}, as well as on the materials involved. However, quite often $E_n$ is considered as depending only on the \textit{initial} neighborhood of the adatom, through a bond-counting approximation. Namely, if $n$ is the number of intra-layer NN atoms of the adatom before hopping, so $E_n = n E_{NN}$, where $E_{NN}>0$ is the NN bond energy. This initial value approximation, usually referred to in the literature as the Clarke-Vvedensky (CV) \cite{CV,*CV1,*CV2} model, obeys detailed balance and has become very popular as a generic and somewhat realistic model for homoepitaxial thin film growth, especially for molecular beam epitaxy (MBE) \cite{barabasi,Krugbook}. In fact, the CV model has been widely employed in the modeling of island nucleation and growth during submonolayer deposition \cite{Ratsch1,Ratsch2,Maria1,tiago13submono}, as well as in the study of multilayer growth \cite{Shitara,Vvedensky1}. Furthermore, different modifications of the CV model have been considered as, e.g, the addition of an Ehrlich-Schwoebel (ES) \cite{Ehrlich,Schwoebel} step-edge barrier, which is key to explain mound formation \cite{Evans2006,Meng,SilvioMound,TiagoMound}.

The kinetic roughening process of surfaces simulated with the CV model (and those from homoepitaxial MBE growth in general) is expected to belong to the Villain-Lai-Das Sarma (VLDS) \cite{Villain,LDS} universality class, aka nonlinear MBE class \cite{barabasi}. Namely, the films' height field $h(\vec{x},t)$ is expected to evolve in the long time limit according to the VLDS equation \cite{Villain,LDS}
\begin{equation}
\frac{\partial h(\vec{x},t)}{\partial t} = -\nu_4 \nabla^4 h + \lambda_4 \nabla^2(\nabla h)^2 + \eta(\vec{x},t),
\label{eqVLDS}
\end{equation}
which describes nonlinear growth processes dominated by surface diffusion, as is the case in CV and typical MBE films. In such equation, $\nu_4$ and $\lambda_4$ are phenomenological parameters, while $\eta$ is a Gaussian white noise. Some renormalization approaches have indeed demonstrated that the dynamic scaling of surfaces from CV model follows VLDS class \cite{Vvedensky2,*Vvedensky3,*Vvedensky4}. Namely, their squared width $w_2$ scales in time, $w_2 \sim t^{2\beta}$, and with the size $l$ of the observation window, $w_2\sim l^{2\alpha}$ (for a given time), with the same exponents $\alpha$ and $\beta$ from the VLDS equation \cite{Janssen}. From a numerical side, however, the situation is more controversial. While some Monte Carlo simulation studies of CV-related models on two-dimensional (2D) substrates --- which is the case relevant for thin film deposition and the single one discussed here --- indicated nonuniversal temperature-dependent exponents \cite{Kotrla}, other works have confirmed VLDS scaling for a broad range of model parameters \cite{Wilby,FabioLADP,ThiagoFabio,TiagoMound}.

A number of recent works have demonstrated, notwithstanding, that universality in surface growth goes far beyond the scaling exponents. Importantly, the (1-point) height distributions (HDs) and (2-point) spatial covariances, which are central quantities determining the statistics of fluctuating surfaces, are believed now to display universal behaviors, beyond an interesting dependence on geometry. In fact, this has been widely established for systems belonging to the Kardar-Parisi-Zhang (KPZ) \cite{KPZ} class in both 1D \cite{Prahofer2000,Sasamoto2010,Amir,Calabrese2011,Takeuchi2010,Takeuchi2011,Alves11,tiago12a,Alves13,HealyCross,silvia17} and 2D \cite{healy12,tiago13,healy13,Ismael14} substrates, whereas a single study exists indicating the same for the VLDS class \cite{Ismael16a}, as well as another one for linear universality classes \cite{Ismael19lin}. In face of this, HDs and spatial covariances have been used in several recent works as a tool for determining the universality class of real thin film surfaces \cite{Almeida14,healy14exp,Almeida15,Yuri15,Almeida17,Rodolfo17}. Despite this appealing application, as far as we know, HDs and covariances have never been investigated for realistic models for thin film growth. For instance, all results for the VLDS class reported in \cite{Ismael16a} were obtained from simulations of ``toy-models'' (with limited mobility of adatoms). Therefore, the aim of the present work is two-fold: (\textit{i}) try to confirm the universality of HDs and covariances for the VLDS class in 2D through the analysis of these quantities for the CV model; and, conversely, (\textit{ii}) investigate the effects of temperature ($T$), deposition flux ($F$) and bond energy strengths on such quantities. As will be demonstrated in the following, the CV HDs display strong and long-living corrections for low $T$ (and/or large $F$), turning almost impossible to firmly establish their universality class in feasible deposition times. At high $T$, however, they agree with that for other VLDS models. The covariances present mild finite-time corrections and always agree with the VLDS one at long time. We simulate also a simplified (limited mobility) model \cite{FabioLADP} which mimics some features of the CV model in the limit of irreversible aggregation, and the same behavior was observed there. 

The remainder of the paper is organized as follows. In Sec. \ref{secModels} the investigated models are defined, as well as the quantities used in the analyses of their surfaces. Results for the HDs and spatial covariances are presented in Secs. \ref{secHDs} and \ref{secCovS}, respectively. Section \ref{secConc} summarizes our final discussions and conclusions.

\section{Models and quantities of interest}
\label{secModels}

The Clarke-Vvedensky (CV) \cite{CV,*CV1,*CV2} model is a version of the classical solid-on-solid (SOS) model by Gilmer and Bennema \cite{Gilmer}, for the regime of complete condensation, when particle desorption from the surface is negligible. The SOS condition implies that overhangs are not allowed at the surface, so that the films formed are compact. In the CV model, particles are deposited at random positions of the substrate with a homogeneous flux $F$ per adsorption site. Here, we consider the substrate as a square lattice of lateral size $L$, with periodic boundary conditions, which is an approximation for films growing in (001) orientation. The growth is performed on flat substrates. Once deposited, an adsorbed particle --- the adatom --- passes to diffuse at the surface through the hopping to nearest-neighbor (NN) sites. The total hopping rate is given by $\mathcal{H} = \nu e^{-(E_d + n E_{NN})/k_B T}$, where $n$ is the number of intra-terrace NN atoms of a given adatom before hopping. The frequency $\nu$ was considered as $\nu = k_B T/(2\pi\hbar)$ in the original CV formulation, as expected from transition theory, with $\hbar$ being the Planck's constant, $k_B$ the Boltzmann's constant and $T$ the temperature. However, it is quite common to consider $\nu$ simply as a constant ($\nu \sim 10^{12}-10^{13} $Hz) \cite{Evans2006,Krugbook} and we will adopt this simpler definition here. In order to model the growth of a specific system with the CV model, we should also specify reliable values for the activation energy barriers $E_d$ and $E_{NN}$ for such a system. In the present work, notwithstanding, we are interested in investigating general properties of the height fluctuations of CV films for broad ranges of $F$, $T$ and energies. So, instead of dealing with all parameters of the model explicitly, it is convenient to rewrite the hopping rate as $\mathcal{H}=D \varepsilon^n$, where $D = \nu e^{-E_d/k_B T}$ is the hopping rate of free adatoms and $\varepsilon = e^{-E_{NN}/k_B T}$ is related to the probability of an adatom ``breaking a bond'' with a lateral NN one. Since the growth is controlled by the ratio $\mathcal{H}/F=(D/F)\varepsilon^n$, it is interesting to define also $R = D/F$, so that $R$ and $\varepsilon$ will be the model parameters. 

We investigate also a limited mobility SOS model introduced by Aar\~ao Reis \cite{FabioLADP}, where again particles are randomly deposited onto a flat square lattice substrate, but only the freshly deposited adatom is allowed to diffuse at the surface. Such adatom can diffuse (by randomly hopping to NN sites, similarly to CV model) while it is free and once it arrives (by deposition or diffusion) at a site with at least one in-plane NN atom, it irreversibly aggregates there. If no one of such sites is found after $G$ diffusion steps, the adatom stops diffusing and permanently aggregates at its final position. Therefore, $G$ is the only parameter in this lateral aggregation of diffusing particles (LADP) model. As demonstrated in \cite{FabioLADP}, by tuning $G$ this model can yield surfaces with the same roughness properties of the CV model with irreversible aggregation, i.e., with $\varepsilon = 0$. Moreover, strong numerical evidence that this model belongs to the VLDS class has been provided in \cite{FabioLADP,FabioLADP2}.

In both CV and LADP models, the time unity will be defined as the deposition of one monolayer (ML) of particles. Here, times up to $t=100000$ will be analyzed, corresponding to the deposition of $10^5$ MLs. We remark that this number of MLs is at least one order of magnitude larger than those considered in typical experiments, as well as in recent numerical works on these and related models \cite{FabioLADP,FabioLADP2,ThiagoFabio}. Results for lateral substrate sizes $L=1024$ (and $L=512$ in some few cases) will be presented below. For each set of model parameters, at least one hundred of different films were grown, totaling $26\times10^6$ points in the statistics. Some less accurate simulations for $L=256$ and $L=2048$ were also performed to confirm that finite-size effects are negligible in our data.

According to the so-called ``KPZ ansatz'', the height at a given point of the films' surfaces, during the transient growth regime, is expected to evolves as \cite{Krug1992,Prahofer2000}
\begin{equation}
h \simeq v_{\infty} t + (\Gamma t)^{\beta} \chi + \ldots,
 \label{ansatz}
\end{equation}
where the growth velocity is always $v_{\infty}=1$ for the models considered here, $\Gamma$ is a model-dependent parameter setting the roughness amplitude, and $\beta$ is the universal growth exponent. Moreover, $\chi$ is a random variable fluctuating according to a probability density function $P(\chi)$, i.e., the underlying height distribution (HD), which is expected to be universal, but dependent on the surface geometry. The films analyzed here are flat and, hence, we will compare our results with previous ones for this geometry. The $n$th centered moment $\expct{h^n}_m$ of the HDs is defined as $\expct{h^n}_m=\left\langle{\overline{\left( h - \bar{h} \right)^n}}\right\rangle$, where the bars indicate averages over the heights of each film surface, at a given time, while $\expct{\cdots}$ denotes the average over different films. In order to characterize the HDs (and other distributions as well) one usually analyzes adimensional ratios of their moments, being the skewness $S=\expct{h^3}_m/\expct{h^2}_m^{3/2}=\expct{\chi^3}_m/\expct{\chi^2}_m^{3/2}$ and the kurtosis $K=\expct{h^4}_m/\expct{h^2}_m^{2}-3=\expct{\chi^4}_m/\expct{\chi^2}_m^{2}-3$ of primary interest.

The two-point spatial statistics of the films' surfaces can be investigated through the correlator
\begin{equation}
 C_S(r,t) = \left\langle \tilde{h}(\vec{x}+\vec{r},t)\tilde{h}(\vec{x},t) \right\rangle,
 \label{eqCov}
\end{equation}
where $\tilde{h}\equiv h - \left\langle h\right\rangle $. From dynamic scaling theory, one expects that $C_S(r,t) \simeq w_2 F(r/\xi)$, where $\xi$ is the correlation length parallel to surface, with $F(x)$ being an, in principle, universal and geometry-dependent spatial covariance. Once it is usually hard to obtain accurate estimates of $\xi$ and the covariances obtained here for the CV model display a pronounced minimum at $r=r_{min}$ (due to a modulated oscillatory decreasing behavior, analogous to that found in other models with dynamics dominated by surface diffusion \cite{Ismael16a,Ismael19lin}), we will analyze the rescaled covariances by plotting $C_S(r,t)/w_2$ against $r/r_{min}$. In this way, all rescaled curves coincide at $r/r_{min}=0$ (where $C_S/w_2=1$) and have minima at $r/r_{min}=1$, so that they shall collapse whenever they have the same shape.

\section{Results for the height distributions}
\label{secHDs}

In this section, we focus on the behavior of the height distributions (HDs) during the transient growth regime. Initially, the case of irreversible aggregation is discussed, which is followed by an analysis of the full CV model.

\subsection{CV model with irreversible aggregation}

Before analyzing the full CV model, it is interesting to start considering the limiting case $\varepsilon = 0$. This corresponds to strong intra-terrace adatom-adatom interactions ($E_{NN} \gg k_B T$), where effectively only free adatoms can move and there is only a single parameter, $R$, to be considered. Although this seems an oversimplification, we notice that this ``irreversible aggregation CV'' (IACV) model has already been investigated for multilayer growth \cite{FabioLADP,Martynec}. Moreover, it is a very common practice to assume that aggregation is irreversible in studies of island nucleation and growth during the submonolayer regime, where the model considered here is a kind of multilayer version of the situation with ``critical nucleus'' $i^*=1$ \cite{Venablesbook,Evans2006,Krugbook,tiago11submono,tiago16PIM,Einax}. 

\begin{figure}[t]
	\includegraphics[width=8.5cm]{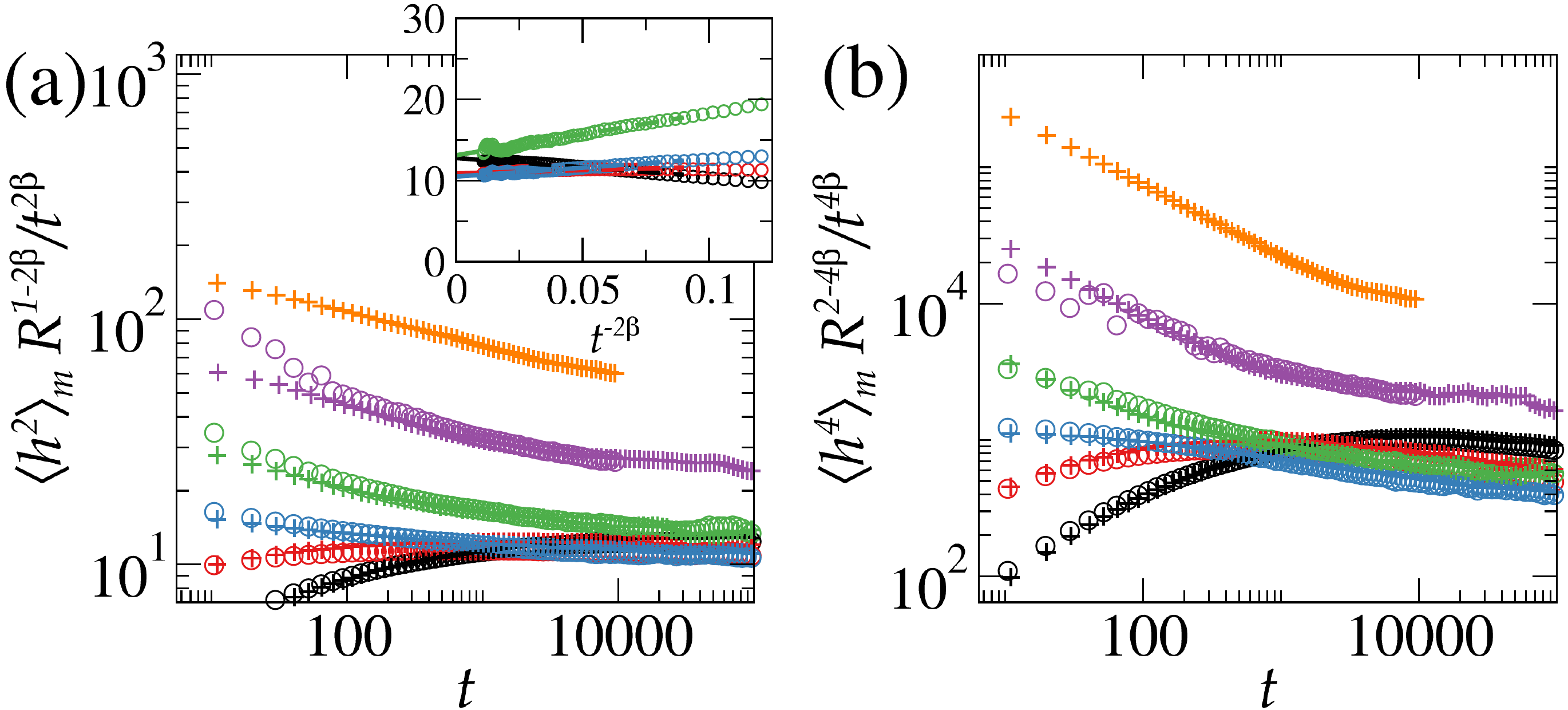}
	\caption{Rescaled HDs' moments $\expct{h^n}_m R^{n/2-n\beta}/t^{n\beta}$ versus time, with (a) $n=2$ and (b) $n=4$, for the IACV (circles) and LADP (plus symbols) models. For the IACV model, data for $R=10$ (black), $10^2$ (red), $10^3$ (blue), $10^4$ (green) and $10^5$ (purple) are shown, while the corresponding parameters for the LADP model are $G=1$ (black), $4$ (red), $15$ (blue), $59$ (green), $222$ (purple) and $846$ (orange). The insertion in (a) shows the same data from the main plot for the IACV model against $t^{-2\beta}$, where the lines are the linear fits used to extrapolate the data.}
	\label{fig1}
\end{figure}

Let us start investigating the effect of $R$ on the amplitude $\Gamma$ in the ``KPZ ansatz'' (Eq. \ref{ansatz}). From such equation, one expects that $\expct{h^n}_m = (\Gamma t)^{n\beta} \expct{\chi^n}_m+\ldots$, for $n\geqslant 2$, and thence the quantity $g_n \equiv \expct{h^n}_m/t^{n\beta}$ should converge to a constant $g_n \rightarrow \Gamma^{n\beta} \expct{\chi^n}_m$ at long times. As demonstrated by Aar\~ao Reis \cite{FabioLADP}, the surface roughness $\left(W=\sqrt{\expct{h^2}_m}\right)$ for the IACV model follows the scaling $W=(L^{\alpha}/R^{1/2}) f(\xi/L)$, with the lateral correlation length $\xi = (R t)^{1/z}$, where $z=\alpha/\beta$ is the dynamic exponent. Since the scaling function $f(x)$ behaves as $f(x) \sim x^{\alpha}$ in the growth regime, we shall have $\expct{h^2}_m \simeq A t^{2\beta}/R^{1-2\beta}$, with $\beta\simeq 0.1975$ being the exponent of the 2D VLDS class \cite{Janssen}. In fact, rescaled curves of $\expct{h^2}_m R^{1-2\beta}/t^{2\beta}$ versus $t$ for not so large $R$ tend to a plateau, where they collapse reasonably well, as shows Fig. \ref{fig1}(a). This confirms the scaling behavior and allow us to estimate the amplitude $A \simeq 12 (2)$ (see the extrapolation in the insertion of Fig. \ref{fig1}). Moreover, this shows that $\Gamma^{2\beta} \expct{\chi^2}_m = A/R^{1-2\beta}$ and, by assuming that $\expct{\chi^2}_m$ is universal (i.e., independent of $R$) one has $\Gamma \sim R^{-\gamma}$, with the exponent $\gamma=1/(2\beta)-1 \approx 1.5316$. If this is the case, rescaled curves of $\expct{h^n}_m [R^{\gamma}/t]^{n\beta}$ for the higher moments should also converge to constant values as $t \rightarrow \infty$. An example of this is shown in Fig. \ref{fig1}(b), for the fourth moment ($n=4$), where one sees that curves for different $R$'s do not collapse so well as those for $n=2$ [in Fig. \ref{fig1}(a)]. It is noteworthy, however, that such rescaled curves do not present clear plateaus even for the smaller $R$'s, showing that the higher moments have stronger finite-time corrections. This possibly explains the absence of a good data collapse and suggests that the plateau will occur for $B \equiv \expct{h^4}_m R^{2-4\beta}/t^{4\beta} \lesssim 500$, when $t \rightarrow \infty$. We notice that this is consistent with the 2D VLDS class, since $B/A^2=\expct{h^4}_m/\expct{h^2}_m^2=K+3$ and a kurtosis $K \approx 0$ was found in \cite{Ismael16a} for the VLDS HDs, so that $B = 3 A^2$ is expected to be found approximately in the range $[300,600]$. For large $R$, we do not observe any plateau or collapse in Fig. \ref{fig1}, certainly because the moments are still far from the VLDS scaling regime [$\expct{h^n}_m \sim t^{0.1975 n}$].

\begin{figure}[t]
	\includegraphics[height=3.65cm]{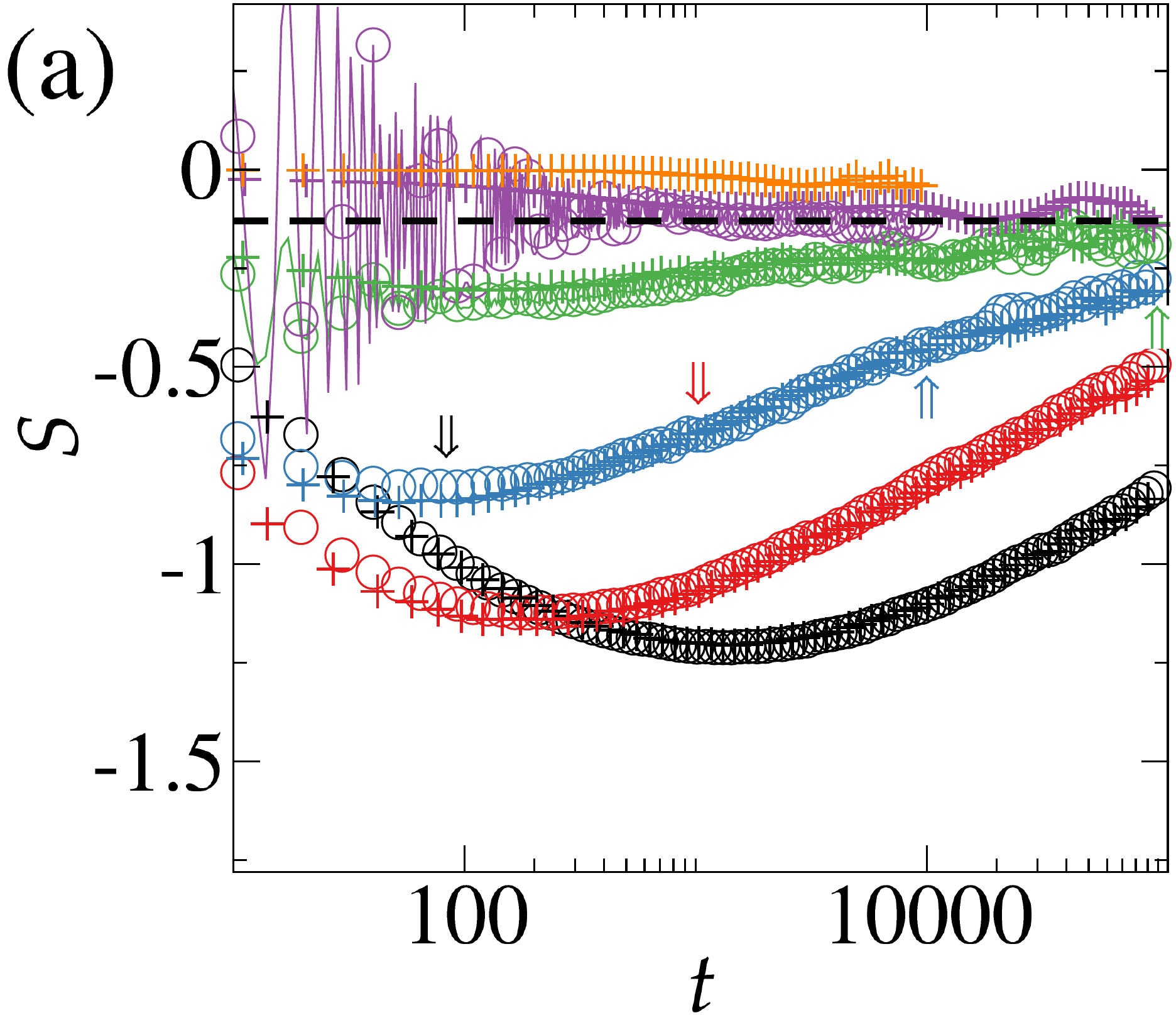}
	\includegraphics[height=3.65cm]{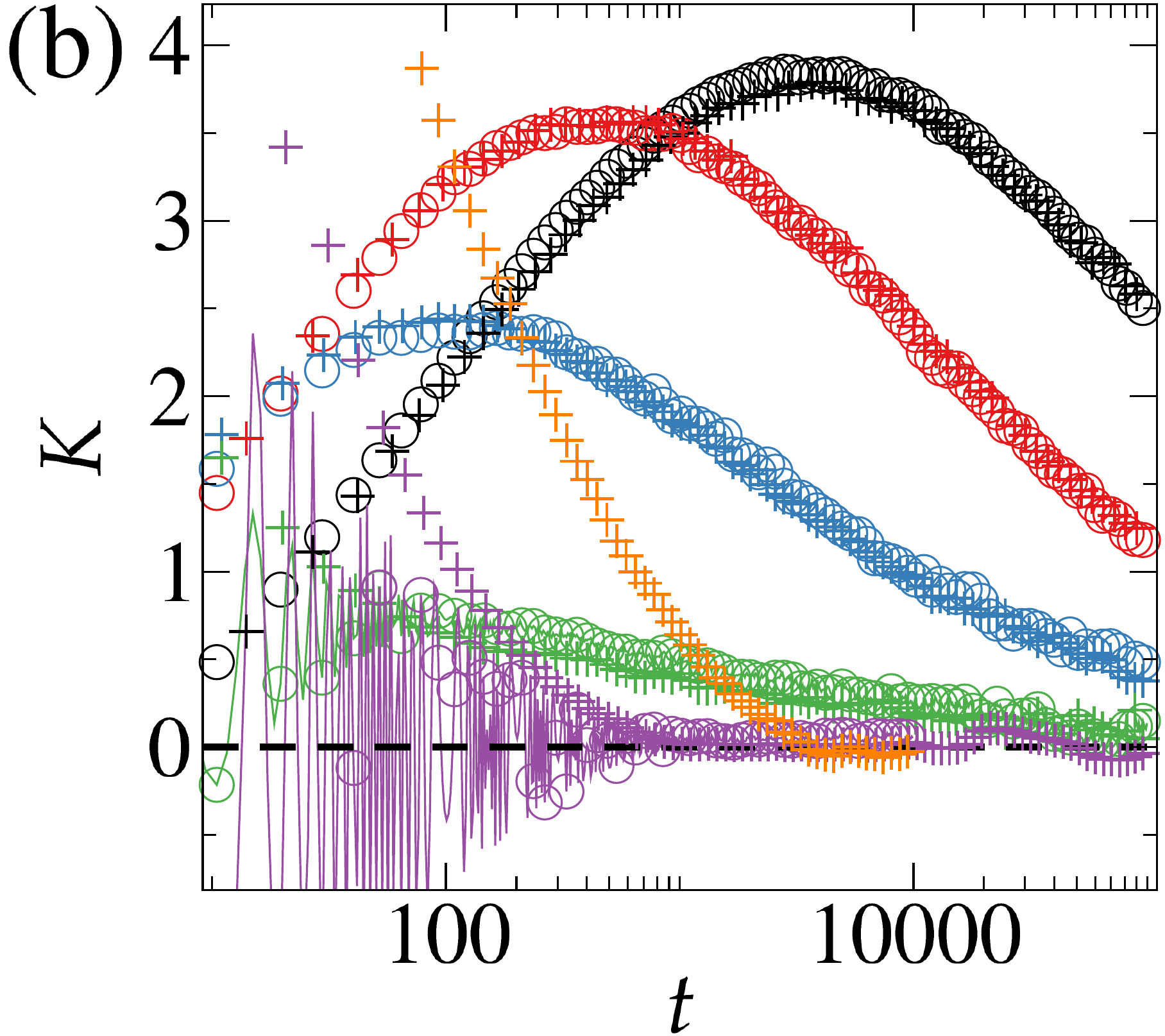}
	\caption{Temporal evolution of the (a) skewness $S$ and (b) kurtosis $K$ for the IACV (circles) and the LADP (plus symbols) models, for the same parameters and color scheme from Fig. \ref{fig1}. The dashed lines represent the universal values of $S$ and $K$ for the 2D VLDS class estimated in Ref. \cite{Ismael16a}. The arrows indicate the times at which the covariances displayed in Fig. \ref{fig6}(a) were measured.}
	\label{fig2}
\end{figure}

The temporal evolution of the skewness $S$ and kurtosis $K$ of the HDs are depicted in Figs. \ref{fig2}(a) and \ref{fig2}(b), respectively. As one can see, these quantities present very strong finite-time effects for small $R$, with very pronounced minima (maxima) in $S$ ($K$), and they seem to be far from their asymptotics even after the deposition of $10^5$ MLs. For $R \gtrsim 10^5$, however, such non-monotonic convergence gives place to an initial oscillatory behavior in $S$ and $K$, due to a layer-by-layer growth at (relatively) short times, and their values converge to the ones numerically estimated for other models belonging to 2D VLDS class in flat geometry: $|S_{VLDS}| \approx 0.13$ and $K_{VLDS}\approx 0.00$ \cite{Ismael16a}. This suggests that for any $R$ the HDs shall agree with the VLDS one, but huge deposition times are needed to observe this when $R$ is not so large. In fact, the ratios for $R=10^4$ also agree reasonably well with $S_{VLDS}$ and $K_{VLDS}$ at long times (see Fig. \ref{fig2}).

As demonstrated in \cite{FabioLADP}, by setting $G \approx 0.28 R^{0.58}$ in the LADP model, it is capable of reproducing the roughness behavior of the IACV model with parameter $R$. Namely, the curves of $W \times t$ for both models collapse. This leads us to inquire whether the same thing happens with the higher moments and their ratios and, quite interestingly, the answer is positive. In fact, rescaled curves of $\expct{h^2}_m$ and $\expct{h^4}_m$ for the LADP model with $G=1$, 4, 15, 59 and 222 - which would mimic the IACV surfaces for $R=10$, $10^2$, $10^3$, $10^4$ and $10^5$, respectively - are also shown in Figs. \ref{fig1}(a) and \ref{fig1}(b) and they remarkably collapse with the IACV ones. A similar agreement is observed also in $S$ and $K$ for both models, for the same parameters, as seen in Fig. \ref{fig2}. This demonstrates that not only the variance of the HDs coincide, but the full HDs are approximately the same for both models, at a given time, for that choice of parameters. Therefore, since simulations of the LADP model are much faster than those for the IACV one, we can use it to investigate the HDs' behavior for larger $R$. For instance, figures \ref{fig1} and \ref{fig2} also show the moments and ratios, respectively, for the LADP model with $G=846$, corresponding to the IACV model with $R=10^6$. We remark that this would be extremely hard to simulate with IACV model for the times, sizes and number of samples we are considering. These results are in consonance with the ones for smaller $R$'s and, particularly, indicate that for $R > 10^5$ the skewness passes to converge to $S_{VLDS}$ from above. The finite-time effect in $K$ is also stronger than that observed in $R = 10^5$. This is indeed expected because by increasing $R$ the duration of the transient layer-by-layer regime also increases. Therefore, although the LADP model does not display such regime, it seems that its surfaces take similar times as the IACV ones to attain the HDs' asymptotic regime.

\subsection{CV model with reversible aggregation}

\begin{figure}[t]
	\includegraphics[width=4.25cm]{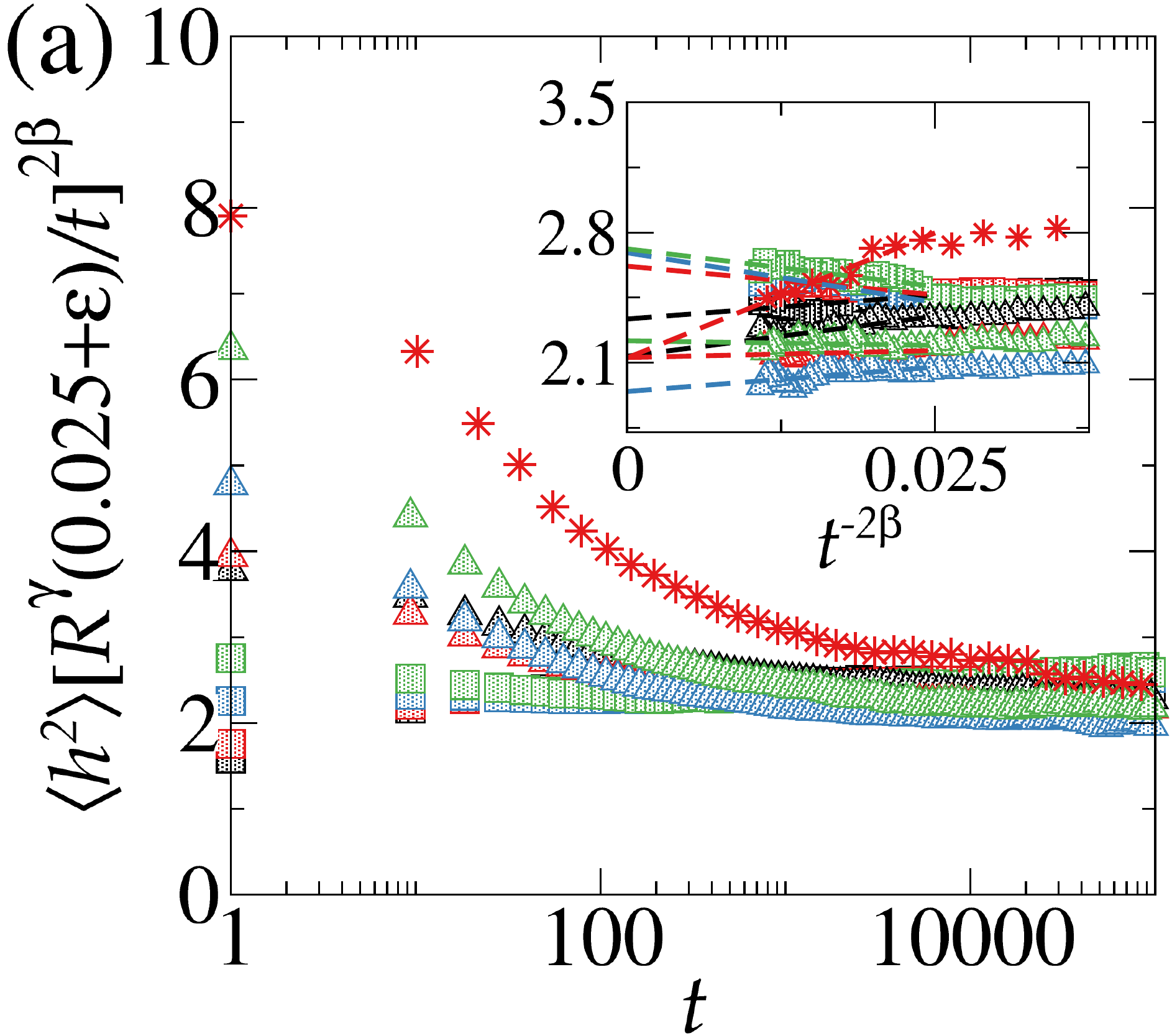}
	\includegraphics[width=4.25cm]{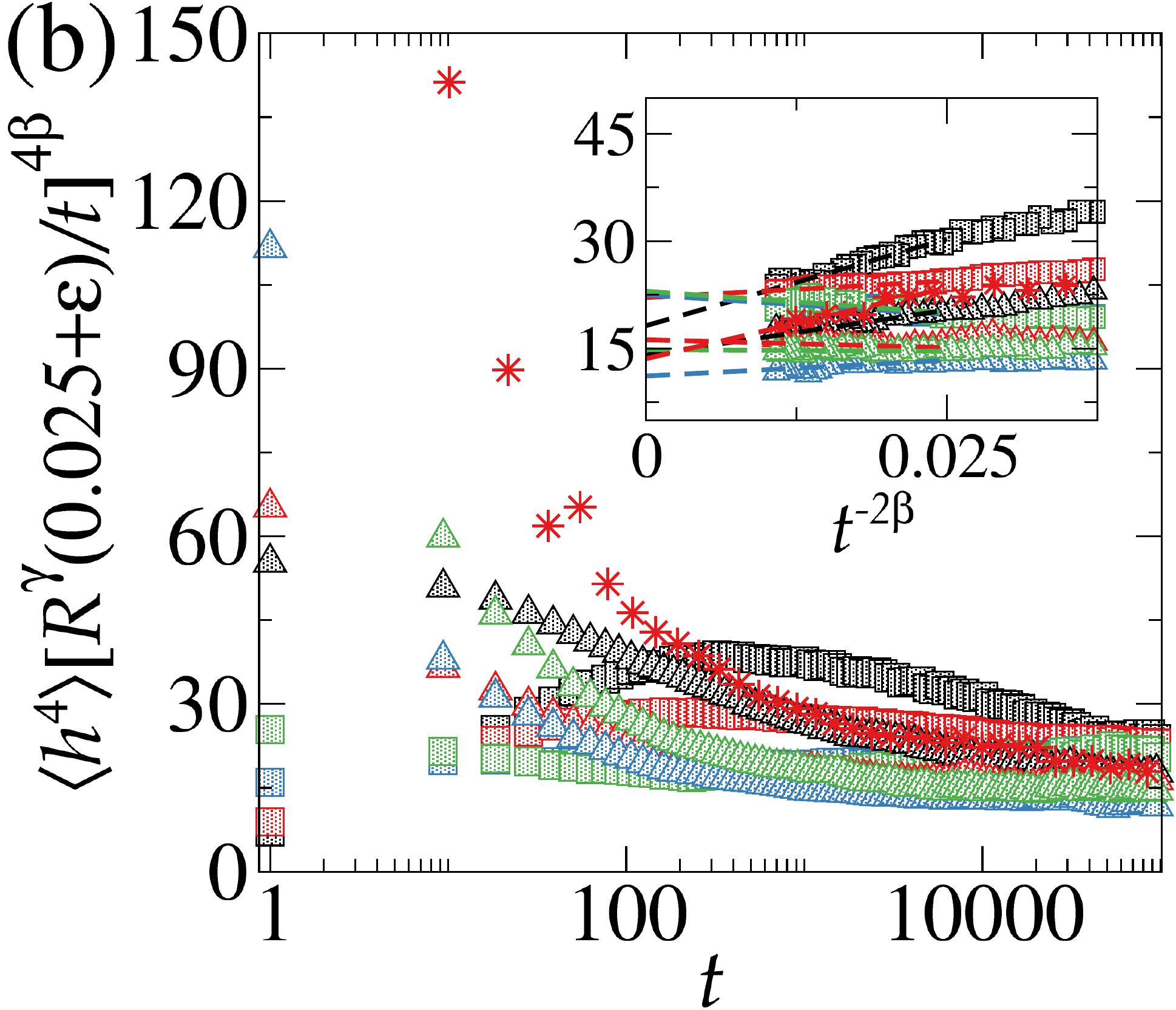}
	\caption{Rescaled HDs' moments $\expct{h^n}_m [R^{\gamma}(0.025+\varepsilon)/t]^{n\beta}$ versus time, for (a) $n=2$ and (b) $n=4$, for the CV model with $R=10^2$ (squares), $10^3$ (triangles), and $10^4$ (stars), and $\epsilon=0.001$ (black), 0.01 (red), 0.05 (blue), and 0.10 (green symbols). The insertions show the same data from the main plots against $t^{-2\beta}$, where the lines are the linear fits used in the extrapolations.}
	\label{fig3}
\end{figure}

Now, we consider the original CV model, with a non-null probability $\varepsilon$ for an adatom detaches from a lateral neighbor. Once again, it is interesting to analyze how the nonuniversal parameter $\Gamma$ in Eq. \ref{ansatz} depends on $R$ and $\varepsilon$. In Ref. \cite{ThiagoFabio}, numerical evidence was provided that during the growth regime the surface roughness behaves as $W \sim \mathcal{F}\left(\frac{t}{R^{\gamma}(0.025+\varepsilon)}\right)$, indicating that $\Gamma \sim \frac{1}{R^{\gamma}(0.025+\varepsilon)}$, where $\gamma= 1/(2\beta)-1$ as in the previous subsection. This means that plots of $\expct{h^n}_m [R^{\gamma}(0.025+\varepsilon)/t]^{n\beta}$ versus $t$ should converge to a single plateau as $t \rightarrow \infty$, what is somewhat confirmed here, as shows Fig. \ref{fig3}. For the second moment a reasonable collapse is found for long times, with the plateau at $A' = 2.5(5)$ [see Fig. \ref{fig3}(a)]. In comparison with the amplitude $A$ above, we have $A'/A = 0.025^{2\beta}\approx 0.23$, which is consistent with our numerical estimate $A'/A = 0.21(7)$. From the discussion in the previous subsection, for the 2D VLDS class, a plateau at $B'= 3A'^2$ approximately in the range [10,30] is expected in the fourth moment, which is indeed observed in Fig. \ref{fig3}(b). This is in agreement with the results above for the IACV model, indicating that the finite-time effects are stronger in the higher-order moments. It is important to remark that the simple dependence of $\Gamma$ on $R$ and $\varepsilon$ presented here, can also present corrections. For instance, rescaled curves for $R=10$ [not shown] do not collapse so well with those depicted in Fig. \ref{fig3}. Moreover, in general, the collapses can be considerably improved by including (somewhat arbitrary) logarithmic corrections in $\Gamma$ [not shown]. Anyhow, a firm conclusion about the existence and form of such corrections would require simulations for larger $R$'s and then larger substrate sizes and much longer deposition times than those considered here, which are not feasible with our current computer resources.

\begin{figure}[!t]
	\includegraphics[width=4.25cm]{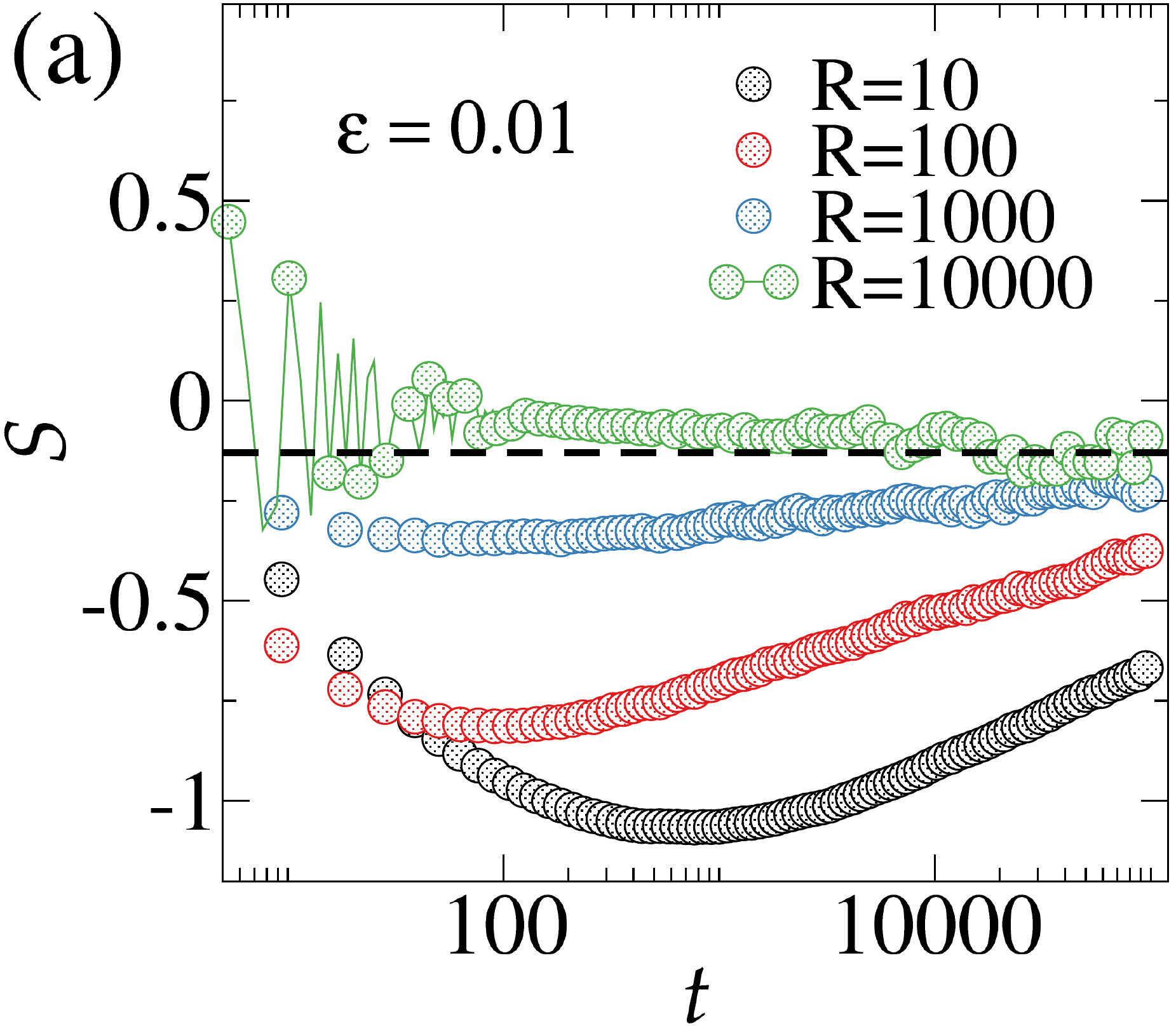}
	\includegraphics[width=4.25cm]{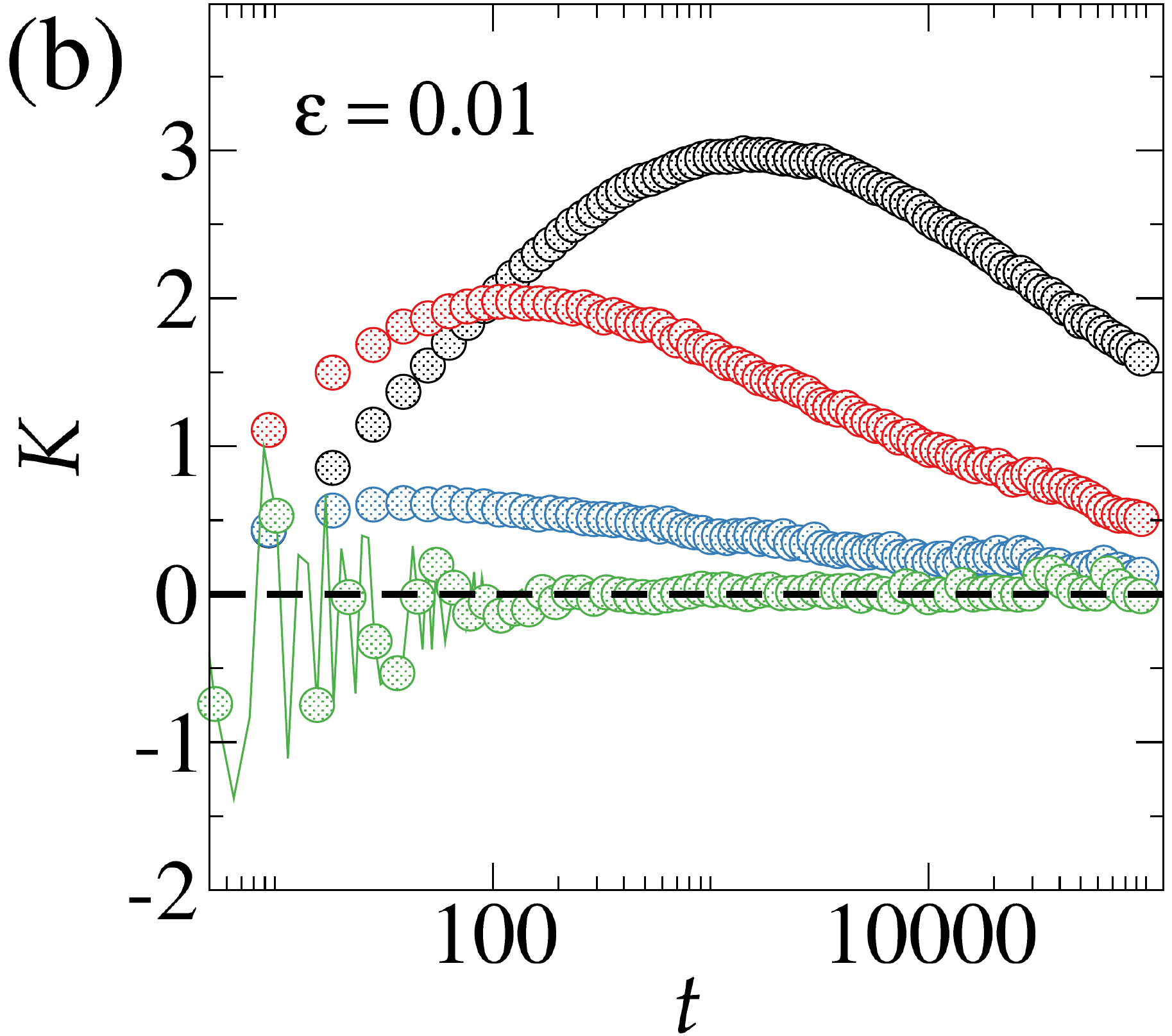}
	\includegraphics[width=4.25cm]{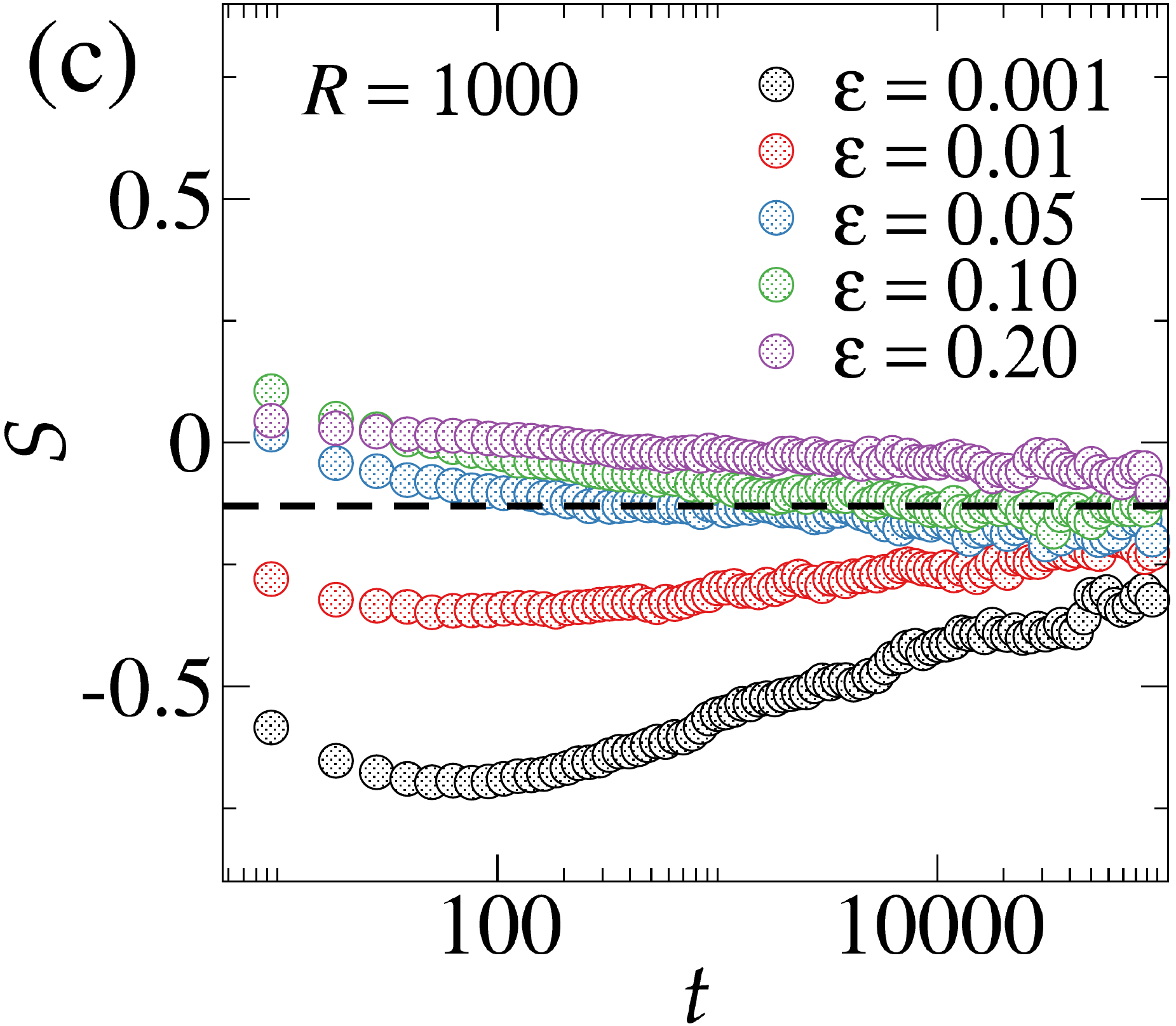}
	\includegraphics[width=4.25cm]{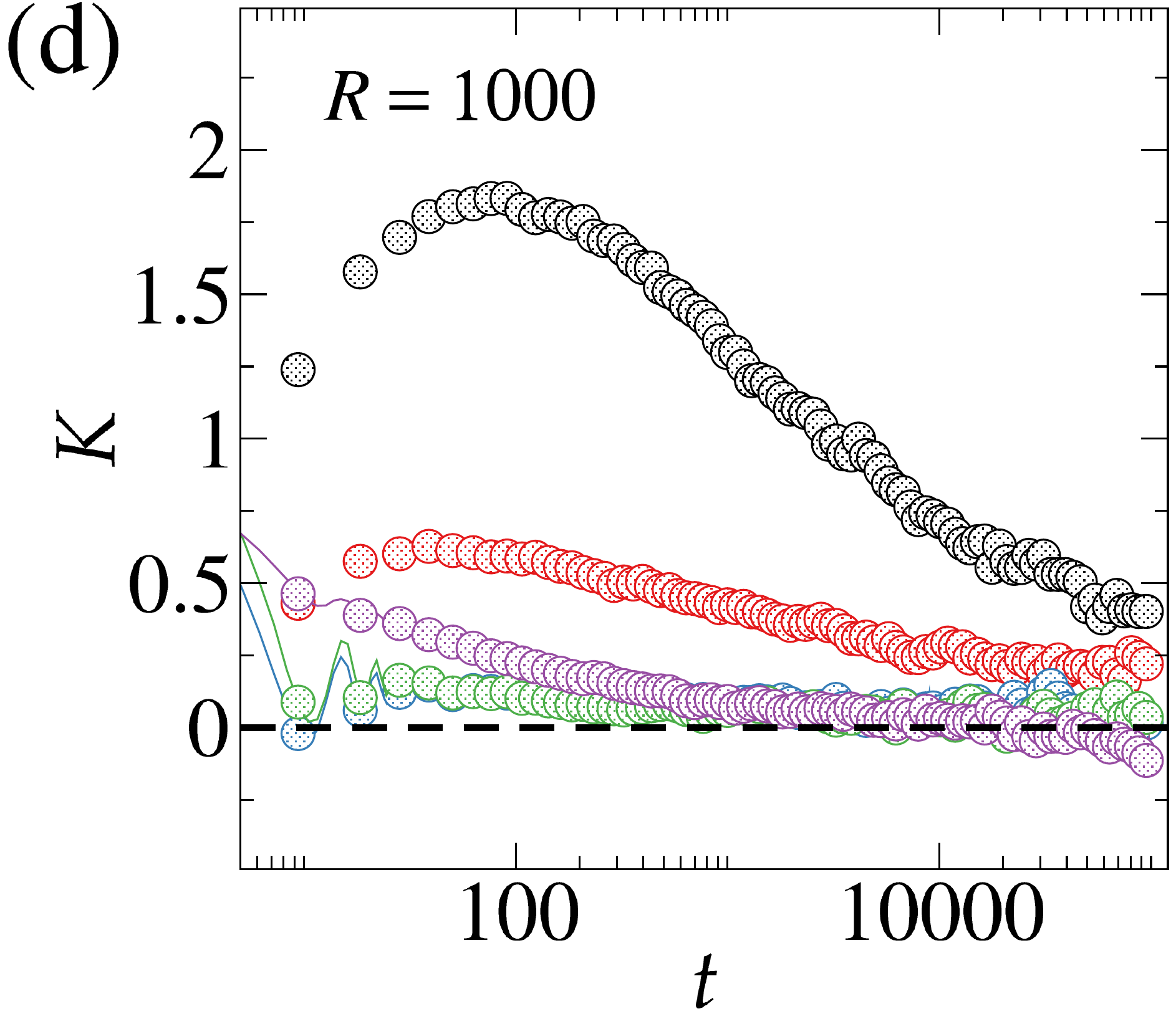}
	\includegraphics[width=4.25cm]{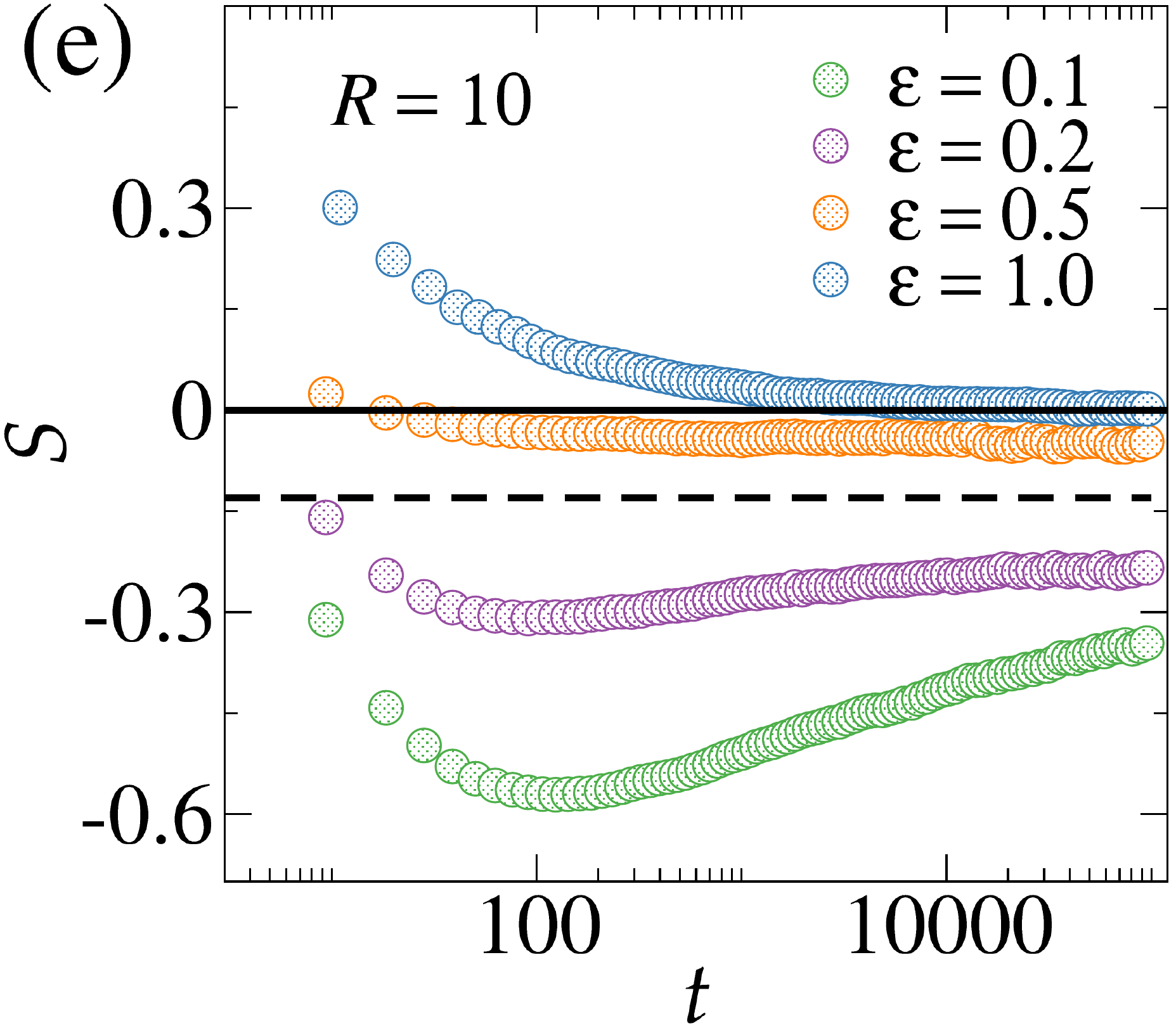}
	\includegraphics[width=4.25cm]{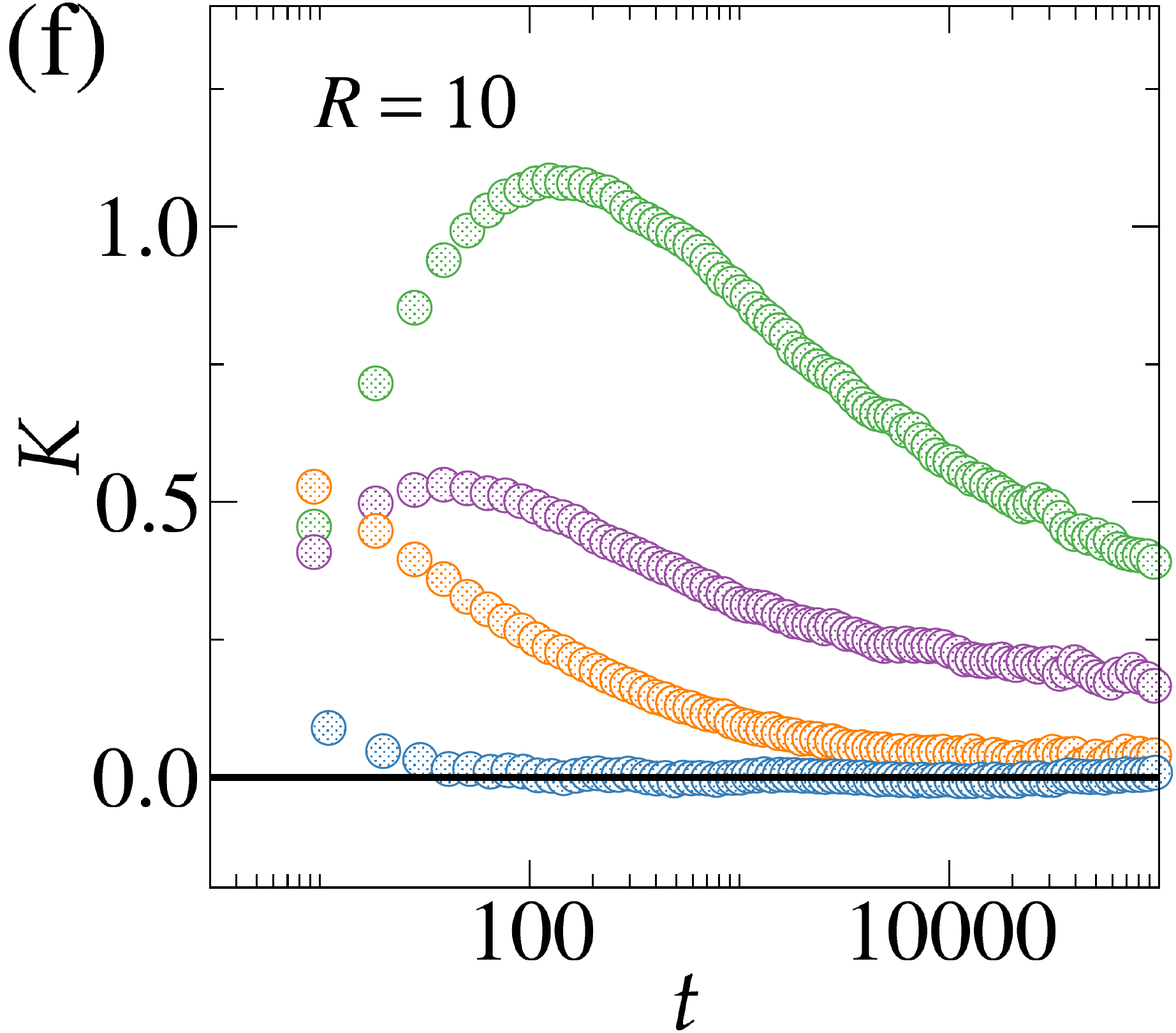}
	\caption{Temporal evolution of $S$ (left) and $K$ (right panels) for the CV model with $\epsilon=0.01$ and several values of $R$ (top), $R=1000$ (middle) and $R=10$ (bottom panels) with several values of $\varepsilon$, as indicated. The dashed lines represent the values $S_{VLDS}$ and $K_{VLDS}$, while the solid lines in (e) and (f) are the cumulant ratios of a Gaussian distribution ($S=K=0$).}
	\label{fig4}
\end{figure}

\begin{figure}[!t]
	\includegraphics[width=8.5cm]{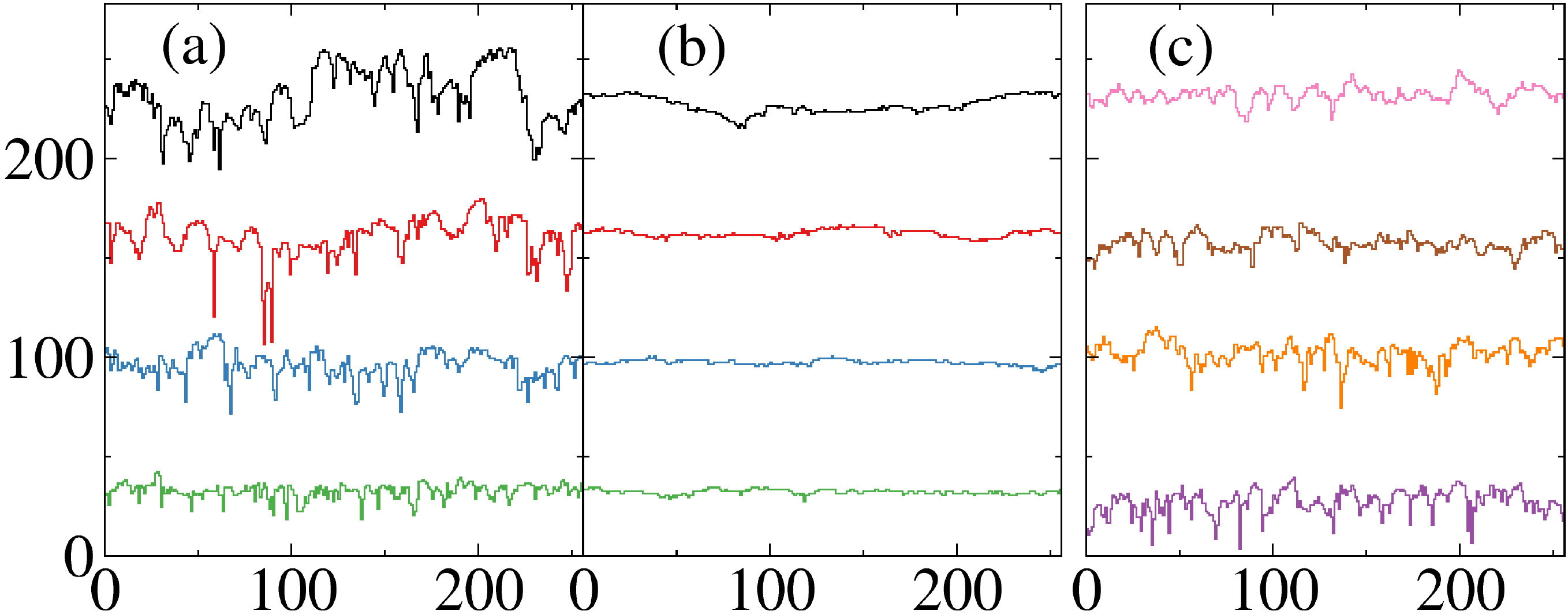}
	\caption{Temporal evolution of 1D cross-sections of $256\times 256$ CV surfaces for (a) $R=10$ and (b) $R=10^3$, both with $\epsilon=0.01$ and times (from the bottom to top) $t=10^2$, $10^3$, $10^4$, and $10^5$. (c) 1D cross-sections of $256\times 256$ surfaces of CV films for $R=10$, $t=1000$ and $\varepsilon=0.001$, $0.01$, $0.1$, and $0.2$  (from the bottom to top). All profiles were shifted vertically to allow their visualization in the same plot and originally they had average heights $\bar{h}\approx t$.}
	\label{fig5}
\end{figure}

Figures \ref{fig4}(a) and \ref{fig4}(b) present the variation in time of the HDs' skewness $S$ and kurtosis $K$ for $\varepsilon = 0.01$ and several values of $R$, where a behavior analogous to that found for the IACV model (Fig. \ref{fig2}) is observed. Namely, for small $R$ severe finite-time effects appear, hallmarked by minima (maxima) in $S$ ($K$), and it is hard to determine whether these ratios converge to the same or different asymptotic values. For the larger $R$'s, on the other hand, there is a faster convergence and the asymptotic values agree quite well with those for the VLDS class. Since the ratio $R=D/F$ determines how much the free adatoms diffuse at the surface before the deposition of a new one, these results and those from the previous subsection demonstrate that when such diffusivity is low the HDs suffer with strong corrections, which decreases as $R$ increases. Note that the low adatom diffusivity for small $R$ yields the nucleation of a large number of 2D islands during the submonolayer regime, over which the subsequent 3D islands grow. Once an adatom is deposited over one of such 3D islands, it has a small probability of escaping from it, for small $R$, so that such islands tend to grow fast vertically. This yields deep valleys between them in the films' surfaces, which certainly explains the rising of HDs with large negative skewness at (relatively) short times. Since the 3D structures tend to form plateaus at their centers, this explains also the large positive kurtosis. As such structures coalesce, however, the deep valleys tend to diminish and then $|S|$ and $K$ start decreasing. By increasing $R$, fewer islands are nucleated in the submonolayer regime, forming fewer valleys, or even a layer-by-layer growth appears at short times for large enough $R$. In any case, this prevents the advent of strong finite-time corrections in the HDs. This scenario is qualitatively confirmed in Figs. \ref{fig5}(a) and \ref{fig5}(b), where the temporal evolution of characteristic 1D cross-sections of films' surfaces are compared for $R=10$ and $R=1000$, respectively. In the former case, we find that surfaces are indeed featured by deep valleys at short times, which tends to decrease for very large $t$, while much more smooth morphologies are found for $R=1000$.

According to the reasoning above, we might expect also to find a decrease in the finite-time effects by increasing $\varepsilon$, for a given $R$, because a large detachment rate of adatoms from steps also contributes to prevent the formation of deep valleys in the films' surfaces. This is indeed observed in Fig. \ref{fig5}(c), where typical 1D profiles for CV films for $R=10$ are shown for the same deposition time ($1000$ MLs) and different values of $\varepsilon$. A quantitative confirmation of this is given in Figs. \ref{fig4}(c) and \ref{fig4}(d), where $S$ and $K$ as function of time are depicted for CV films with $R=10^3$ and several values of $\varepsilon$. There, one sees that the minima (maxima) in $S$ ($K$) curves decrease as $\varepsilon$ increases and for $\varepsilon \gtrsim 0.05$ they converge to the VLDS ratios. This strongly suggests that for the smaller $\varepsilon$'s they shall also converge to there, but taking much longer times.

Results for $R=10$, however, indicates that this is not the case for large values of $\varepsilon$. In fact, curves of $S$ for $\varepsilon \gtrsim 0.10$ converge to nonuniversal $\varepsilon$-dependent values at long times [see \ref{fig4}(e)]. We remark that this is somewhat expected because the case $\varepsilon=1$, where adatoms do not interact with their NN ones (since $E_{NN}=0$), corresponds to a kind random deposition. As a matter of fact, although adatoms are still diffusing, they only stop moving when they are earthed by another adatom, due to a random deposition or a random diffusion towards it. These random processes alone do not generate correlations in the system and, thus, the surface roughness increases as $W \sim t^{1/2}$  and the HDs are Gaussian. This is indeed confirmed in Fig. \ref{fig4} (c), where the skewness for $\varepsilon=1$ converges to $S=0$. Moreover, the kurtosis for this parameter also converges to $K=0$ [see Fig. \ref{fig4}(f)]. The same is true for larger values of $R$. We can see in Fig. \ref{fig4}(c) that the curve of $S$ for $R=1000$ and $\epsilon=0.2$ seems do not converge to $S_{VLDS}$ or it will take a very long time to arrive at that.

\section{Results for the spatial covariances}
\label{secCovS}

\begin{figure}[t]
	\includegraphics[width=4.cm]{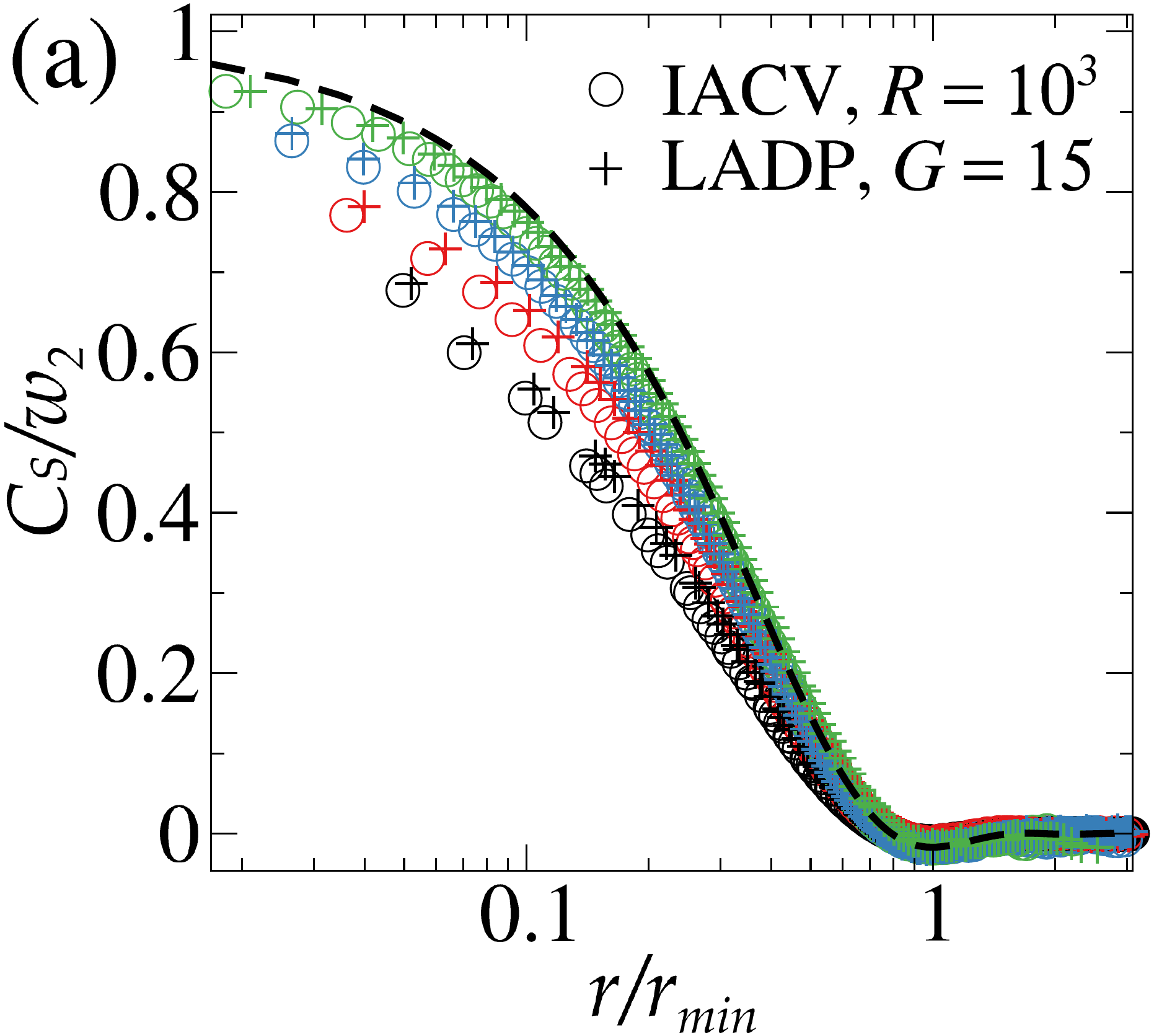}
	\includegraphics[width=4.cm]{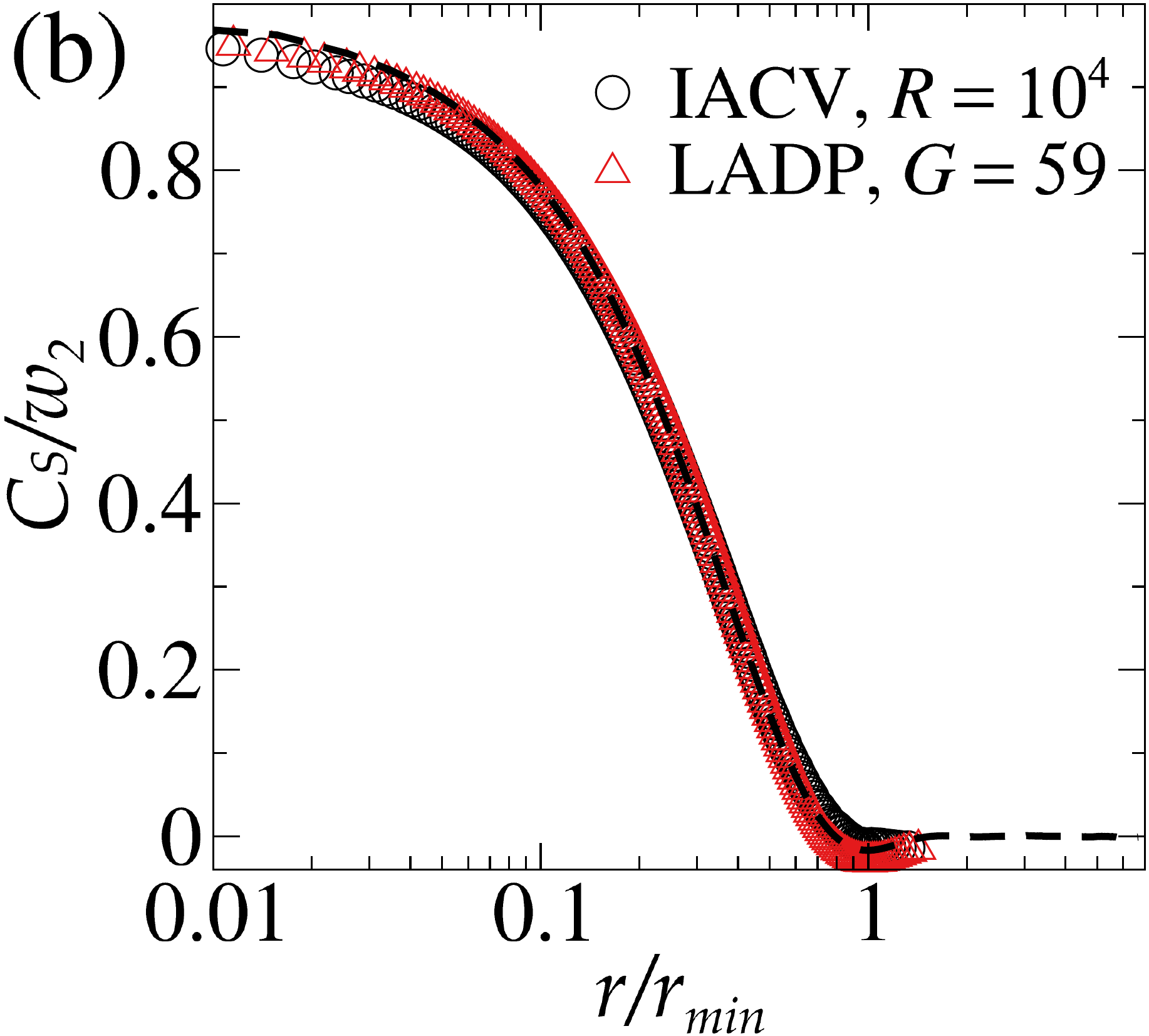}
	\includegraphics[width=4.cm]{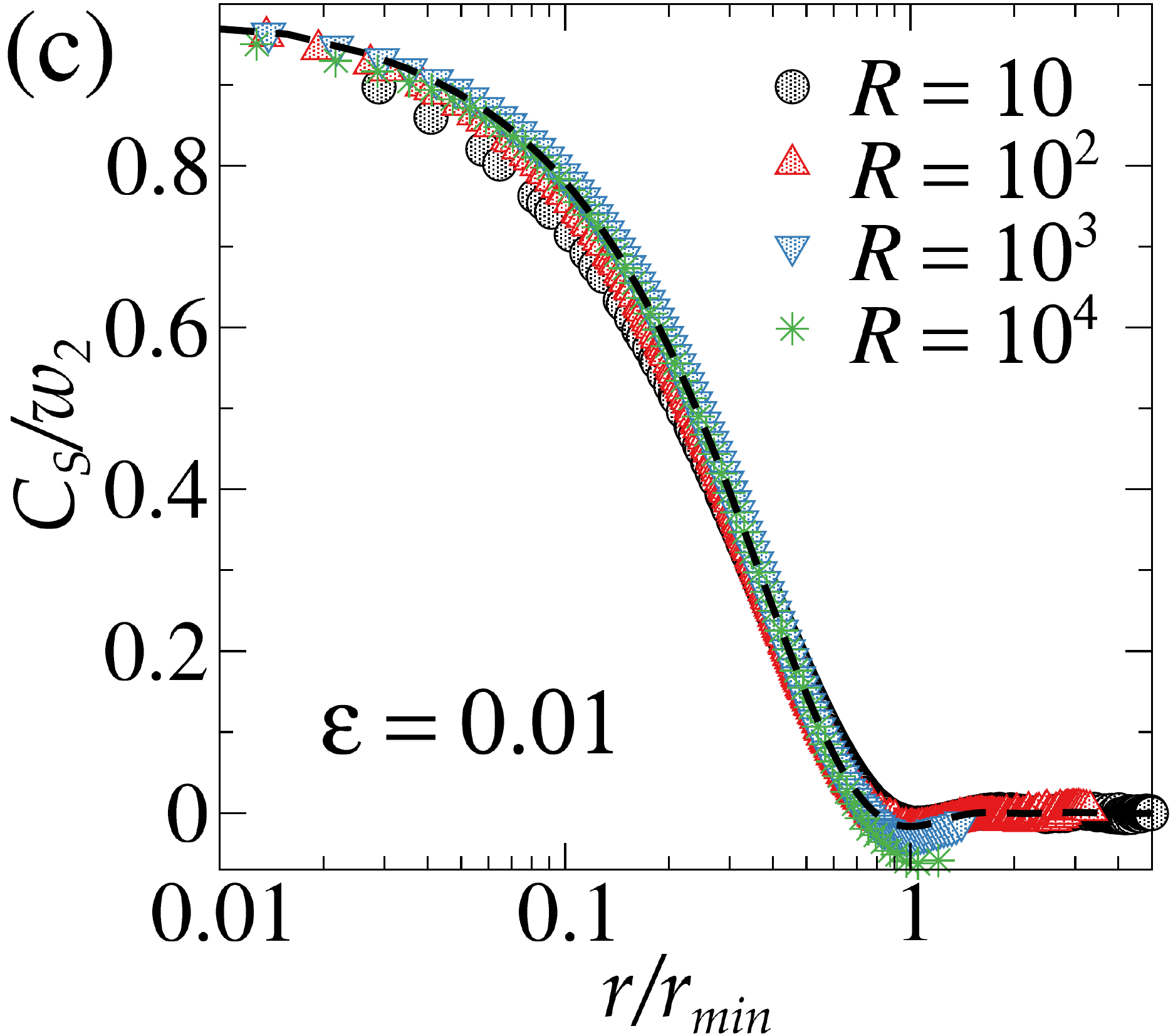}
	\includegraphics[width=4.cm]{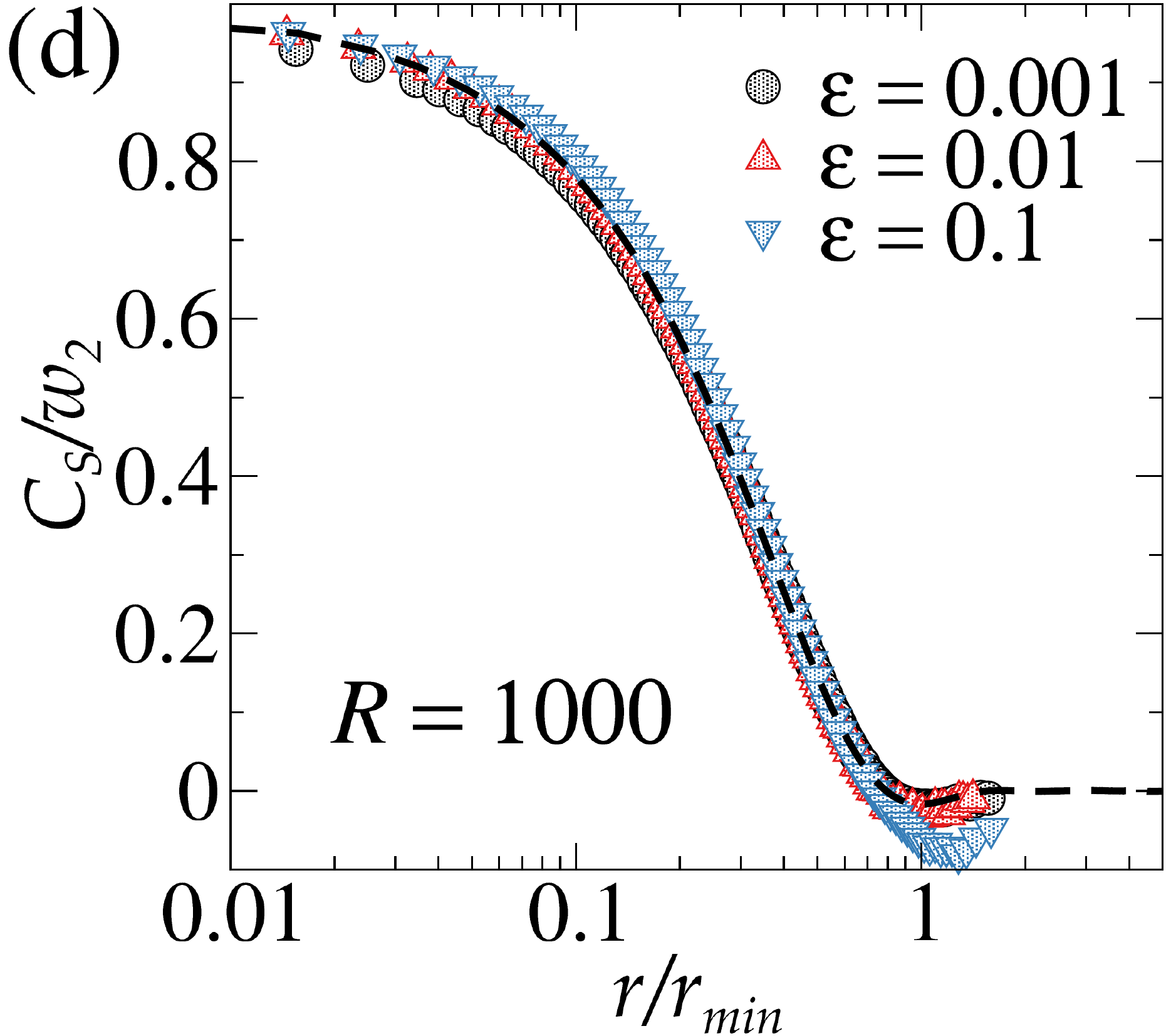}
	\caption{Rescaled spatial covariances $C_S/w_2$ versus rescaled length $r/r_{min}$. (a) Comparison of curves for the IACV (circles) and LADP (plus symbols) models, for the times $t=10^2$ (black), $10^3$ (red), $10^4$ (blue), and $10^5$ (green), as indicated by the arrows in Fig. \ref{fig2}(a). Data for the same models, at $t=10^4$, but with parameters yielding a larger adatom diffusivity are shown in panel (b). (c) Curves for the CV model with $\epsilon=0.01$ with several $R$'s. (d) Curves for the CV model with $R=1000$ with several $\epsilon$'s. All data in (c) and (d) are for $t=10^5$. In all panels, the dashed line is the universal covariance curve for the 2D VLDS class, estimated in \cite{Ismael16a}.}
	\label{fig6}
\end{figure}

Now, we discuss the two-point spatial covariance function (Eq. \ref{eqCov}) of CV films' surfaces. Once again, let us start with the case of irreversible aggregation. Figure \ref{fig6}(a) shows rescaled covariance curves for the IACV model with $R=1000$ and the LADP one with $G=15$, for the times indicated by the arrows in Fig. \ref{fig2}(a). We observe that for short times, when the HDs are quite different from the VLDS one --- i.e., the values of $S$ and $K$ are far from $S_{VLDS}$ and $K_{VLDS}$ [see Figs. \ref{fig2}(a) and (b)] --- the covariances are also different from the VLDS curve, which was numerically estimated by us in \cite{Ismael16a}. However, as time increases, the covariances converge towards the VLDS one, presenting a reasonable collapse with it for the longest time considered here. This makes clear that the same sort of finite-time effects observed in the HDs are also present in the covariances. Figure \ref{fig6}(b) presents data for the IACV model with $R=10^4$ and its LADP counterpart --- for which the HDs agree well with the VLDS one already for $t \gtrsim 10^4$ --- and, in fact, a quite good collapse of the rescaled covariances with the VLDS one is observed for $t=10^4$. Therefore, similarly to what happens in the HDs, as the adatom diffusivity increases the finite-time corrections decrease, uncovering the VLDS universality in the system. It is important to emphasize also the good agreement between the IACV and LADP covariances for the parameters $G \approx 0.28 R^{0.58}$. This indicates that the entire statistics of IACV films' surfaces are captured by the simplified LADP model. 

The results for the CV model are shown in Figs. \ref{fig6}(c) and \ref{fig6}(d) for $\varepsilon=0.01$ and several $R$'s, and $R=10^3$ and several $\varepsilon$'s, respectively. In this case, the time dependence is similar to that just discussed for the irreversible aggregation [Fig. \ref{fig6}(a)], so, only results for the longest simulation time are presented in Figs. \ref{fig6}(c) and \ref{fig6}(d). There, one can see that in general the curves for the CV model with very different parameters collapse quite well among them, as well as with the VLDS curve. Especially for $R=10$, although a slight deviation is observed in Fig. \ref{fig6}(c) for small $r/r_m$, the overall behavior is quite close to the VLDS one. This demonstrates that the finite-time corrections in the covariances are milder than those found in the HDs. Furthermore, the agreement with the VLDS covariances provides a strong confirmation that the CV model belongs to VLDS class.

\section{Conclusions}
\label{secConc}

We have presented an extensive analysis of the 1-point and 2-point statistics of surfaces of the CV model for homoepitaxial thin film growth, for broad ranges of model parameters ($R\in [10,10^5]$ and $\varepsilon \in [0,1]$), covering a wide range of temperature (and/or deposition flux) and energy strengths. For instance, if one assumes that $\nu=10^{13}$ s$^{-1}$ and $E_d = 1$ eV, which are close to the values expected in the growth of several semiconductor, metal and organic films \cite{Evans2006,Krugbook}, for a flux $F=1$ ML/s we have simulated temperatures approximately in the interval $T = 420-630$ K. Conversely, for a typical temperature value, e.g., $T = 420$ K, one has the flux in the interval $F = 0.0001-1.0$ ML/s. In any case, these parameters are consistent with those commonly used in actual thin film deposition experiments  \cite{Evans2006,Krugbook}.

We have also simulated the simplified LADP model \cite{FabioLADP}, where the mobility of adatoms is limited to the freshly deposited one and found that, by appropriately tuning the parameter in such model, it is able to produce surfaces with the same HDs and covariances of the CV model with irreversible aggregation (IACV). This is a rather interesting finding, once the already known fact that both models have the same (actually, very similar) roughness evolution, as demonstrated in \cite{FabioLADP}, does not necessarily imply that their surfaces would have the same 1-point and 2-point statistics. Our results, however, demonstrate that this is indeed the case. As pointed out above and stressed recently in Refs. \cite{Martynec,Tung}, the existence of this kind of simplified model reproducing the surface features of a more realistic and complex one (the IACV model here) is very important since this allows us to investigate regimes of the latter model, which would not be computationally accessible in a feasible amount of time, through simulations of the former one.

Most of our results indicate that the HDs and covariances asymptotically agree with those obtained in Ref. \cite{Ismael16a} in simulations of simplified models belonging to the 2D VLDS class. Substantially, this provides a strong confirmation that the CV model indeed belongs to the VLDS class and, conversely, confirms the universality of the HD and covariance for this class.

It turns out, however, that severe finite-time effects are observed in these quantities, especially in the HDs, for small $R$ (corresponding to low $T$ and/or high $F$). Namely, when the adatom mobility is low when compared with deposition; what leads to the formation of deep valleys in the films' surfaces, which yield the deviations. For a given value of $R$, the atom-atom energy $E_{NN}$ plays also an important role in this matter, enhancing the finite-time effects when it is large (i.e., when $\varepsilon$ is small). Interestingly, this behavior is contrary to the one found in the surface roughness, $W$, where VLDS scaling is observed already at short times for small $R$, but the time it takes to appear increases with $R$, due to the initial layer-by-layer regime which appears for large $R$. For this reason, a clear scaling $W \sim t^{0.1975}$ is not observed in simulations of the CV model for large $R$, even for the deposition of $10^5$ MLs, as shown in Figs. \ref{fig1}(a) and \ref{fig3}(a). Similar conclusions have been reported in \cite{FabioLADP,FabioLADP2,ThiagoFabio}. Hence, HDs and covariances are complementary measures to $W$ in such systems, with the former (later) working better for large (small) $R$'s. For this reason, it is very important to investigate all these quantities together in order to determine the universality class of a given growing film. Moreover, a careful finite-time analysis is imperative in face of the strong effects observed here, which may be present also in other realistic models, as well as in experiments on homoepitaxial thin film growth.

\acknowledgments

This work was supported by CNPq, CAPES, FAPEMIG and FAPERJ (Brazilian agencies).

\bibliography{bibHDsCV}

\end{document}